\documentstyle[12pt,a4]{article}

\begin{document}

\def\case#1#2{\hbox{$\frac{#1}{#2}$}}
\def\slantfrac#1#2{\hbox{$\,^#1\!/_#2$}}
\def\reference#1{\relax\par\noindent}

\def\refZjnl#1{{\rm#1}}
\def\aj{\refZjnl{AJ}}			
\def\araa{\refZjnl{Ann. Rev. Astron. Astrophys.}}		
\def\apj{\refZjnl{Astrophys. J.}}
\def\apjl{\refZjnl{Astrophys. J. Lett.}}
\def\apjs{\refZjnl{ApJS}}		
\def\aap{\refZjnl{A\&A}}		
\def\mnras{\refZjnl{Mon. Not. R. Astron. Soc.}}
\def\prb{\refZjnl{Phys.Rev.B}}		
\def\deg{\hbox{$^\circ$}}
\def\sun{\hbox{$\odot$}}
\def\earth{\hbox{$\oplus$}}

\title{Curvature pressure in a cosmology with a tired-light redshift\\
{\small Australian J. Phys., 1999, {\bf 52}, 753}}

\author{David F. Crawford\\
School of Physics, A28, University of Sydney, N.S.W. 2006, Australia\\
d.crawford@physics.usyd.edu.au}

\maketitle

\begin{abstract}

A hypothesis is presented that electromagnetic forces that prevent ions from following geodesics 
results in a curvature pressure in a Tired-Light Cosmology's. 
It  may partly explain the solar neutrino deficiency and it may be the engine that drives 
astrophysical jets. 
However the  most important consequence  is that, 
using general relativity without a cosmological constant, 
it leads to a static and stable cosmology.
Combined with an earlier hypothesis of a gravitational interaction of photons and particles 
with curved spacetime a static cosmology is developed that predicts a Hubble constant of  
$H=60.2\,\mbox{km}\cdot \mbox{s}^{-1}\cdot \mbox{Mpc}^{-1}$and  a microwave background 
radiation with a temperature of  3.0\,K.  
The background X-ray radiation is explained and observations of the 
quasar luminosity function and the angular distribution of radio sources 
have a better fit 
with this cosmology than they do with standard big-bang models.  
Although recent results (Pahre et al 1996) for the Tolman surface brightness test
favor the standard big-bang cosmology they are not completely inconsistent
with a static tired-light model.
Most observations that imply the existence of dark matter measure redshifts, 
interpret them as velocities, and invoke the virial theorem to predict masses 
that are much greater than those deduced from luminosity's.
If however most of these redshifts are due to the gravitational interaction in  
intervening clouds  no dark matter is required.  
Observations of quasar absorption lines. a microwave background temperature at
 a redshift of $z=1.9731$, type 1a supernova light curves and the Butcher-Oemler
effect are discussed. 
The evidence is not strong enough to completely eliminate a non-evolving cosmology.
The result is a static and stable cosmological model that agrees with most of the current observations.
\end{abstract}

\section{Introduction}
North (1995) provides an excellent history of the discovery that distant
galaxies had a distant-dependent redshift and of the various theories that 
were proposed to explain this redshift.
Although there was strong initial support for tired-light models the lack
of a viable physical explanation and the apparent success of the expansion
cosmology has meant that there has been little consideration of tired-light
models in the last forty years.
However in a short note  Grote Reber (1982) argues that Hubble himself was never a 
promoter of the expanding universe model and personally thought that a 
tired-light model was simpler and less irrational.
LaViolette (1986) has compared the generic tired-light model with the big-bang 
model on four kinds of cosmological tests. 
He concluded that  non-evolving Euclidean tired-light model is a better fit 
for the cosmological
tests of angular diameter verses redshift, magnitude verses redshift, number density of 
galaxies verses magnitude and number density of radios sources verses flux density. He also 
provides  references to earlier theories and to his own model for a tired-light mechanism.

The strongest theoretical arguments against a tired light model are that it requires new physics 
and that any scattering mechanism that gives rise to an energy loss will also produce an angular 
scattering that is not observed. 
In the tired-light model considered here the new physics for the redshift is minimal 
and the angular scattering is insignificant. 
In previous papers (Crawford 1987a, 1991) it was argued  that there is a gravitational 
interaction  such that photons and particles  lose energy as they pass through a gas. 
The energy loss for photons' results in a redshift that could produce the Hubble redshift.
The argument is that photons can be treated as discrete entities with
a finite extent that are subject to the "focusing" theorem of curved spacetime. 
That is the cross-section of a bundle of geodesics (in a space with positive curvature) 
will decrease in area with distance. 
This is just the analogue of the convergence of lines of longitude. 
The hypothesis (Crawford 1987a) is that this focusing produces an interaction that leads to the
loss of energy to very low energy secondary photons and an effective redshift of the primary 
photon.
Because the interaction is effectively with the mass that produces the curved spacetime and 
that this mass will have a very much larger inertia than the particle the angular 
scattering of the photon will be negligible.

If the Hubble redshift is explained by a non-expansion mechanism there is still
the problem that a static cosmological solution to the equations of general relativity
is unstable so that a viable cosmology requires some way to provide stability.
This problem was investigated (Crawford 1993) without curvature pressure in a
static cosmology within a Newtonian context. 
That work is superseded by the present paper which shows that, using general
relativity, the use of curvature pressure can provide a static and stable
cosmology.
The solution is the new concept of curvature pressure which  is based on the observation  
that in plasmas electromagnetic forces completely dominate 
the particle motions  so that they do not travel along geodesics. 
The curvature pressure is the reaction back on the material that generates curved 
spacetime from the non-geodesic motion of its component particles.
Where curved spacetime  is due to a plasma the reaction  is seen as a (curvature) pressure 
within the plasma that depends on its density and temperature and acts to prevent compression.  

The curvature pressure is investigated in a static cosmological model and for plasmas that 
occur in the center of the sun and around compact objects. It is shown that the effect of 
curvature pressure will  decrease the central solar temperature by an amount that may be  
sufficient to explain the observed deficiency of solar neutrinos.  
Since curvature pressure acts to oppose contraction and since it increases with temperature 
it is unlikely that black holes could form from hot plasmas. However it remains  possible to 
form black holes from cold material. More significantly curvature pressure is very important 
in accretion disks around compact objects and may provide the engine that drives 
astrophysical jets. 

Since the big-bang cosmological model in all its ramifications is so well entrenched,  
to be taken seriously,  any alternative model must at least be able to explain  
the major cosmological observations.
It is argued that using the Friedmann equations the introduction of curvature pressure leads 
to a static and stable cosmological model. 
One of the predictions of this model  is that there is a background X-ray radiation and an 
analysis of the background observations done in a previous paper are used to determine the 
average density of the universe. 
Because of its essential importance to this static cosmology and because the earlier results 
did not include the effects of curvature pressure the hypothesis of a gravitational 
interaction is revisited. 
The result is a prediction  of the Hubble constant 
$H=60.2\,\mbox{km}\cdot \mbox{s}^{-1}\cdot \mbox{Mpc}^{-1}$.  
and the existence of the microwave background radiation with a temperature of  3.0\,K. 
It is  shown how the observations that lead to the occurrence of dark matter in the big-bang 
cosmology are readily explained without dark matter.
Next previous work on the luminosity function of quasars and the angular sizes of radio 
sources is discussed to show that the observations can be fitted without evolution. 
The classic Tolman surface brightness test is discussed with respect to recent
observations from Pahre et al (1996).
The theme of evolution, or lack of it, is continued with examination of observations on 
quasar absorption lines, a  microwave background temperature at high redshift, 
type 1a supernovae light curves and the Butcher-Oemler effect.
Finally the topics of nuclear abundance, entropy, and Olber's paradox are briefly covered.  

\section{Theoretical background}
A theme that is common to the development of both curvature pressure and the gravitational 
interaction is that in four-dimensional-space the effects of centripetal acceleration are 
essentially  the same as they are in three-dimensional-space. 
Mathematically a smooth three-dimensional curved space can be locally approximated  as a 
four-dimensional Euclidean space.  
Provided the volume is small enough and the curvature is smooth enough 
higher order spaces can be neglected.  
The hypothesis made here is that this four-dimensional space has a physical reality.  
Note that is not to be confused with four-dimensional spacetime; here we have 
five-dimensional spacetime.
Consider two meridians of longitude at the equator  with a perpendicular separation of  $h$, 
then  as we move along the longitudes this separation obeys  the differential equation   
$h''=-h/r^2$ where the primes denote differentiation with 
respect to the path length and  $r$ is  the radius of the earth. 
In addition the particle has a centripetal acceleration of   $v^2/r$ where  $r$  can be 
determined from the behavior of  $h''$. 
In four-dimensional-space the longitudes become a geodesic bundle and the separation  
becomes a cross-sectional area, $A$, where  $A''=-A/r^2$.
Again the particle has a centripetal acceleration of  $v^2/r$ where now  $r$ is the radius 
of the hyper-sphere. Although the particle as we know it is  confined to three dimensions 
there is a centripetal acceleration due to curvature in the fourth  dimension that could 
have significant effects.  
Another fundamental topic considered is the nature  of gravitational force.
It is critical to the development of curvature pressure that gravitation  produces 
accelerations and not forces.

The gravitational interaction theory explicitly requires that photons and particles are 
described  by localized wave packets. 
The wave equations that describe their motion in flat spacetime are  carried over to 
curved spacetime in which the rays coincide with geodesics. 
In particular with the  focusing theorem (Misner, Thorne \& Wheeler 1973)  there is an 
actual focusing of the wave  packet in that its cross-sectional area decreases as the 
particle (photon)  travels along its trajectory.  
In this and previous papers (Crawford 1987a, 1991) it is argued that the result is a 
gravitational  interaction in which the particle  loses energy.

\section{The theoretical model for curvature pressure}
In a plasma there are strong, long-range electromagnetic forces that completely dominate 
accelerations due to gravitational curvature. 
 The result is that, especially for electrons, the 
particles do not travel along geodesics. 
 If we stand on the surface of the earth our natural geodesic  is one of free fall but 
the contact forces of the ground balance the gravitational acceleration with  the 
consequence that there is a reaction force back on the ground. 
The result of stopping our  geodesic motion is to produce a force that compresses the ground.
 The major hypothesis of this  paper is that there is a similar reaction force in 
four-dimensional spacetime. 
This force acts   back on the plasma (that produces the curved  spacetime)  because 
its particles do not follow geodesics. 
Thus the plasma appears in two roles. 
The first produces the curved spacetime and in the second the failure of its particles to 
follow geodesics  causes a reaction back on itself acting in the first role. 
 It is long-range electromagnetic forces that are important, not particle collisions. 
For example in a gas without long-range forces and assuming that the time spent during 
collisions is negligible  the particles will still travel along geodesics between collisions. 
Given that there are long range forces that dominate the particle trajectories there is a  
reaction force that appears as a pressure, the curvature pressure.

For the cosmological model consider  the plasma to occupy the surface of a 
four-dimensional hypersphere.  
It is easier to imagine if one of the normal dimensions is suppressed then it will appear as 
the two-dimensional surface of a three-dimensional sphere. 
The nature of this pressure can then be understood by analyzing this reduced model with 
Newtonian physics in three-dimensional space.
 In this case the curvature pressure acts within the two-dimensional surface and is another 
way of describing the effects of the centripetal accelerations of the particles. 
By symmetry the gravitational attraction on one particle due to the rest is equivalent to 
having the total mass at the center of the sphere.
 To start let the shell contain identical particles all with the same velocity, 
and let this sphere have  a radius r,  then the radial acceleration of a particle with 
velocity v is $v^2/r$. 
At equilibrium the radial accelerations are balanced by the mutual gravitational attraction.
 Now for a small change in radius, $dr$, without any change in the particle velocities 
and going from one equilibrium position to another  we can equate the work done by the 
curvature pressure to the work done by the force required to overcome the centripetal 
acceleration to get
 \begin{equation}
\label{e1}
p_cdA=-{{Mv^2} \over {r}}dr,
\end{equation}
 where M is the total mass, but for a two-dimensional area  $dA/dr=2A/r$ therefore  
$p_c=-Mv^2/2Ar=-\rho v^2/2$ where $\rho $ is the surface density.  
Thus the effects of the centripetal accelerations can be represented as a negative 
pressure acting within the shell. 
The next step is to generalize this result to many types of particles where each 
type of particle has a distribution of velocities. 
 
The particles are constrained  to stay in the shell by a dimensional constraint that is 
not a force.  
The experiments of E\"{o}tv\"{o}s and others (Roll et al. 1964 and Braginski and Panov 1971) 
show that the Newtonian passive gravitational mass is identical to the inertial mass to 
about one part in  $10^{12}$. 
The logical conclusion is that Newtonian gravitation produces an acceleration and not a force. 
The mass is only introduced for consistency with Newton's second law of motion. 
The concept of gravitation as an acceleration and not a force is even stronger in general 
relativity. 
Here the geodesics are the same for all particles independent of their mass and gravitational 
motion does not use the concept of force. 
Clearly for a single type of particle the averaging over velocities is straightforward so 
that the curvature pressure is $p_c=-\rho \overline {v^2}/2$. 
The averaging over particles with different masses is more ambiguous. 
Traditionally we would weight the squared velocities by their masses; that is we 
compute the average energy. 
However since the constraint that holds the particles within the two-dimensional shell is not 
due to forces and since gravitation  produces accelerations and not forces the appropriate 
average is over their accelerations.  
The result for our simple Newtonian model is  
\begin{equation}
\label{e2}
p_c=-\case{1}{2}\rho \sum\limits_i {\overline {v_i^2}},
\end{equation}
where  the density is defined as   $\rho =\sum\limits_i {n_i}m_i$ and $n_i$ is the number 
density of the i'th type of particle. 
This simple Newtonian  model gives a guide to what the curvature pressure would be for a 
more general model in a homogeneous isotropic three-dimensional gas that forms the surface 
of a four-dimensional hyper-sphere. 
The dimensional change requires that we replace  $dA/dr$  by $dV/dr=V/3r$, and  then 
including the relativistic corrections (a factor of  $\gamma ^2$) needed to transform the 
accelerations from the particle's reference system to a common system where the average 
velocity is zero, we get
\begin{eqnarray}
\label{e3}
p_c& = &-{{\rho}  \over {3}}\sum\limits_i {n_i}\overline {\gamma _i^2v_i^2}\nonumber\\ \nonumber
& =& -{{\rho c^2} \over {3}}\sum\limits_i {n_i\left( {\overline {\gamma _i^2}-1} \right)}\nonumber\\
& =&  -{{\rho c^2} \over {3}}{\left( \overline {\gamma^2}-1 \right)},
\end{eqnarray}
 where  the Lorentz factor $\gamma ^2=1/\sqrt {1-v^2/c^2}$. 
 Note that although the equation for curvature pressure does not explicitly include the 
spacetime curvature the derivation requires that it is not zero. 
Because this equation was only obtained by a plausibility argument we hypothesize that 
the curvature pressure in the cosmological model is given by 
equation (\ref{e3}).

Since  the particles may have relativistic velocities, and assuming thermodynamic equilibrium,  
the   ($\overline {\gamma ^2}-1$)  factor can be evaluated using the J\"{u}ttner distribution.  
For a gas at temperature  $T$ and particles with  mass  $m$  de Groot, Leeuwen \& van Weert (1980)  show that 
 \begin{equation}
\label{e4}
\gamma ^2(\alpha )=3\alpha K_3(1/\alpha)/K_2(1/\alpha)+1,
\end{equation} 
where $\alpha = kT/mc^2$ and  $K_n(1/\alpha)$ are the modified Bessel functions of the 
second kind (Abramowitz and Stegun 1968). 
For small  $\alpha $ this has the approximation
\begin{equation}
\label{e5}
\gamma ^2(\alpha )=1+3\alpha +{\case{15}{2}}\alpha ^2+{\case{45}{8}}\alpha ^3+\ldots .
\end{equation}
Note for a Maxwellian distribution the first three terms are exact so that the extra 
terms are corrections required for the J\"{u}ttner distribution. 
For non-relativistic velocities equation (\ref{e5}) can be used and equation (\ref{e3}) 
becomes
\begin{equation}
\label{e6}
p_c=-\frac{1}{N}\sum\limits_{i=1}^N {\left( {{{n_i} \over {m_i}}}\right)}\overline mkT,
\end{equation}
where  $n_i$ is the number density for the i'th type of particle and   
$\overline m=\sum\limits_{i=1}^N {n_im_i}/n$ is the mean particle mass. 
Except for the inverse mass weighting  and the sign this is identical to the expression 
for the thermodynamic pressure.
 
\section{Solar interior  and local plasma concentrations}
The equation for curvature pressure derived above for the cosmological model cannot be 
used in other situations with different metrics. 
The key to understanding the application of curvature pressure in other metrics such as 
the Schwartzschild metric used for stellar interiors is to consider the case where the 
overall curvature is small and superposition  may be assumed. 
Since the free fall acceleration of a particle is independent of its mass there is no 
curvature pressure associated  with external gravitational fields provided they have scale 
lengths much greater than the typical ion separation.  
Any curvature pressure is due to local curvature of the metric produced by the local density.  
This arises because although the electrons and ions have in general different centripetal 
accelerations these are completely dominated by accelerations due to the electromagnetic forces. 
Let the gravitational potential be   $\Phi $, then the self-gravitational energy density 
is $\rho\Phi $. 
Now it was argued above that the curvature pressure is proportional to the energy density 
(it has the same units) but with an averaging over accelerations rather than forces that 
results in replacing   $\rho$ by  $\left( {\overline {\gamma ^2}-1} \right)\rho $. 
Consequently we take the curvature pressure in a plasma due to its own density as
 \begin{equation}
\label{e7}
p_c=\case{1}{3}\left( {\overline {\gamma ^2}-1} \right)\rho \Phi 
\end{equation}  
Note that the derivation is essentially one based on 
dimensional analysis and therefore the numerical factor of $1/3$  may need modification. 
It was used in part for consistency with the cosmological curvature pressure and in part 
because it makes the application of equation (\ref{e7}) to a low temperature gas with a 
single type of particle have the simple expression   $p_c=p_T\Phi /c^2$ where   $p_T$ 
is the thermodynamic pressure.
From potential theory we get for the curvature pressure of a plasma at the point $r_0$ 
the expression
\begin{equation}
\label{e8}
p_c\left( {r_0} \right)=
 \case{1}{3}G\rho \left( {r_0} \right)\left( {\overline {\gamma 
^2\left( {r_0} \right)}-1} \right)
\int \frac{\rho \left( {r-r_0} \right)}{\left| {r-r_0} \right|} dV.
\end{equation}
Equation (\ref{e8}) can be simplified for non-relativistic velocities by using the 
approximation (equation \ref{e5}) to get 
 \begin{equation}
\label{e9}
p_c={{G\rho \left( {r_0} \right)kT} \over {c^2}}\left( {\sum\limits_{i=1}^N 
{{{n_i} \over {nm_i}}}} \right)
\int \frac{\rho \left( {r-r_0} \right)}{\left| {r-r_0} \right|} dV
\end{equation} 
where n is the total number density. 

The curvature pressure adds to the thermodynamic pressure (and radiation pressure) to 
support the solar atmosphere against its own gravitational attraction. 
That is for the same gravitational attraction the required thermodynamic pressure, 
and hence the temperature, will be reduced by curvature pressure. 
Applying equation (\ref{e9}) to the sun and using pressures, temperatures, and abundance 
ratios given by Bahcall  (1989), it was found that the curvature pressure at the center 
of the sun is  $2.8\times 10^{14}\,\mbox{Pa}$ compared to the thermodynamic pressure of   
$2.3\times 10^{16}\,\mbox{Pa}$. 
Since the temperature is directly proportional to the thermodynamic pressure this implies 
that the temperature at the center of the sun is  reduced by 1.2\%.
Bahcall  (1989 p151) shows that the $^8$B neutrino flux is very  sensitive to the temperatures 
at the center of the sun with a flux rate that is proportional to the eighteenth power of 
the temperature. 
Thus this temperature change would decrease the neutrino flux to 80\% of  that from the 
standard model. 
Although the observed ratio of   $2.55/9.5=27\% $ (Bahcall 1997)  is much smaller the effect 
of the pressure curvature is clearly significant and large enough to warrant a more 
sophisticated computation.

\section{Cosmology with curvature pressure}
The main application of curvature pressure is to a  cosmological model  for a homogeneous 
and isotropic distribution of a fully ionized gas. 
Based on the theory of general relativity and using the Robertson-Walker metric the 
Friedmann equations (Weinberg 1972) are
\begin{eqnarray*}
\label{e10}
-\ddot R & =& {{4\pi G} \over {c^2}}\left( {\rho c^2+3p}\right)R\\
 R\ddot R+2\dot R^2 & = & {{4\pi G} \over {c^2}}\left( {\rho c^2-p} 
\right)R^2 -2kc^2,
\end{eqnarray*}  
where R is the radius $\rho $ is the proper density,  $p$ is the pressure, $G$ is the 
Newtonian gravitational constant, and  $c$ is the velocity of light. 
The constant  $k$ is one for a closed universe, minus one for an open universe and zero 
for a universe with zero curvature.  
Working to order  $m_e/m_H$ the thermodynamic pressure can be neglected but not the 
curvature pressure.  
The equations including the curvature pressure (equation \ref{e3}) are
\begin{eqnarray*}
\label{e12}
-\ddot R & = & 4\pi G\rho R\{ 1-\left( {\overline {\gamma ^2}-1} \right)\} \\
R\ddot R+2\dot R^2 & =&  4\pi G\rho R^2\{ 1+\case{1}{3}\left( {\overline {\gamma^2}-1} 
\right)\} \\ & & - 2kc^2, 
\end{eqnarray*}
where  $\overline {\gamma ^2}$ is the average over all velocities and particle types.  
Clearly 
 $\ddot R$ is zero if   $\overline {\gamma ^2}=2$ and equation (\ref{e4})  can be solved 
for a hydrogen plasma to get   $\alpha _e=kT_0/m_ec^2=0.335$ or   $T_0=1.99\times 10^9$K. 
Thus with thermal equilibrium the second derivative of  $R$ is zero if the plasma has 
this temperature.  
Thus the requirement for stability leads to a prediction of the plasma temperature.
This temperature is based on a model in which the plasma is homogeneous, but the occurrence 
of galaxies and clusters of galaxies show that it is far from homogeneous. 
In order to investigate the effects of inhomogeneity consider a simple and quite arbitrary 
model where the plasma is clumped with the probability of a clump having the density n is 
given by the exponential distribution  $\exp \left( {-n/n_0} \right)/n_0$,  where  $n_0$  
is the average density. 
Assuming pressure equilibrium so that  $T_e=T_0n_0/n$ then for  $\overline {\gamma ^2}=2$ 
we find that  the average temperature $T=1.1\times 10^9$K thus showing that the effect of 
inhomogeneity  could reduce the observed temperature
by a  factor of  order two. 

Since the right hand side of the second Friedmann equation is positive then the curvature 
constant $k$ must be greater or equal to zero. 
The only useful static solution requires that   $k=1$ and with  $\dot R=\ddot R=0$ the 
result for the radius of the universe is given by
 \begin{equation}
\label{e14}
{{1} \over {R_0^2}}={{8\pi G\rho _0} \over {3c^2}}.
\end{equation}
Thus the model is a static cosmology with positive curvature. 
Although the geometry is the 
same as the original Einstein static model  this cosmology  differs in that it does not 
require a cosmological constant. 
Furthermore it is stable. Consider a perturbation,  $\Delta R$,  about the equilibrium 
position then the perturbation equation is
 \begin{equation}
\label{e15}
\Delta \ddot R={{c^2} \over {8\pi R_0}}\left( {{{d\overline {\gamma 
^2}} \over {dR}}} \right)\Delta R,
\end{equation}
and since for any realistic equation of state the average velocity  (temperature) will 
decrease as R increases the right hand side is negative showing that the result of a 
perturbation is an oscillation about the equilibrium value. 
Thus this model does not suffer from the deficiency that the static Einstein model has of 
gross instability.  
Since the volume of the three-dimensional surface of the hyper-sphere is  $2\pi ^2R_0^3$ 
the radius of the universe can be written  in terms of the total mass of the universe, 
$M_0$, as
 \begin{equation}
\label{e16}
R_0={{4GM_0} \over {3\pi c^2}},
\end{equation}
which differs by a factor of $2/3$ from that (Crawford 1993) which was derived from a 
purely Newtonian model. 
For interest the values with a density of  $2.05m_H \mbox{ m}^{-3}$ (see below) are  
$R_0=2.17\times 10^{26\,}\mbox{m}\ =7.04\,\mbox{Gpc}$, and  
$M_0=6.90\times 10^{53}\,\mbox{kg}\ =3.47\times 10^{23}\,\mbox{M}_{\mbox{sun}}$.

\section{Background X-ray radiation }
If this cosmological model is correct there should be a very hot plasma between the 
galaxies and in particular between galactic clusters. 
This plasma should produce a diffuse background X-ray radiation and indeed such 
radiation  is observed.  
Attempts to explain the X-rays by bremsstrahlung radiation within the standard 
model have not been very successful (Fabian \& Barcons 1992), mainly because it must 
have originated at earlier epochs when the density was considerably larger 
than present. 
The hard X-rays could come from discrete sources but if it did there are problems
with the spectral smoothness and strong evolution is required to achieve the
observed flux density (Fabian \& Barcons 1992).
However there is an excellent fit to the data in a static cosmology (Crawford 1987b, 1993) 
for X-ray energies between 5\,KeV and 200\,KeV. 
Using universal abundances (Allen 1976) the analysis showed a temperature of 
$1.11\times 10^9$\,K and a density of  $2.05m_H \mbox{ m}^{-3}$.  
Comparison of this temperature with that predicted by the homogeneous model of
$171\,\mbox{KeV}$ shows that it is nearly a factor of two too small.  
A possible explanation comes from the observation that the universe is not homogeneous. 
Although there is fortuitous agreement with the simple inhomogeneous model described above 
this can only be interpreted as showing that the observations are consistent with an 
inhomogeneous model.  

One of the main arguments against the explanation that the background X-ray radiation 
comes from a hot inter-cluster plasma is that this plasma would distort the cosmic 
microwave background radiation  by the Sunyaev-Zel'dovich effect. 
This distortion is usually expressed by the dimensionless parameter $y$.  
Mather et al (1994) have measured the spectrum of the cosmic microwave background 
radiation and conclude that  $\left| y \right|<2.5\times 10^{-5}$. 
In the big-bang cosmology most of the distortion occurs 
at earlier epochs where the predicted density and the temperature of the plasma are 
much higher than current values. 
However for any  static model we can use a constant density of 
 $2.05m_H \mbox{ m}^{-3}$ in the equation (Peebles 1993)
 \begin{equation}
\label{e17}
y={{kT_e\sigma _Tn_er} \over {m_ec^2}},
\end{equation}
where  $\sigma _T$ is the Thomson cross-section and $r$  is the path length since the 
formation  of the radiation. 
For a hydrogen plasma we get  $y=2.6\times 10^{-29}r$. 
The microwave background radiation (see below) it is being 
continuously replenished by energy losses from the hot electrons and the  typical path 
length for the energy lost by electrons to equal the energy of a photon at the peak of 
the spectrum is  
 $3.5\times 10^{18}\,$m which results in  $y=9.1\times 10^{-11}$ well within the observed limits.

\section{The Hubble constant}
One  of the major requirements of any cosmological model is the necessity to explain the 
relationship found by Hubble that the redshift of extra-galactic objects depends on 
their distance.  
In earlier papers (Crawford 1987a, 1991) the author suggested that there is an interaction 
of photons with curved spacetime that  produces an energy loss that can explain the 
Hubble redshift relationship.  
Because the earlier work did not include the effects of curvature pressure and because 
this interaction is central to the description of a viable static cosmology a brief  
updated description is given here. 
The principle is that a photon can be considered as a localized wave traveling along a 
geodesic bundle. 
Because of the `focusing theorem' (Misner et al 1973) the cross-sectional area of  this 
bundle will decrease with time, and in applying this theorem to a photon it was argued 
that this will cause a change in the photon's properties. 
In particular angular momentum will decrease because it is proportional to a spatial 
integral over the cross-sectional area.  
The change in angular momentum can only be sustained for a time consistent with the 
Heisenberg uncertainty principle. 
The conclusion is that eventually there will the emission of two (in order to conserve 
the total angular momentum)  very low energy photons.

 The second part of the argument is that the rate at which this energy loss occurs is 
proportional to the rate of change of area of the geodesic bundle.  
This rate of change of area in the absence of shear and vorticity is given by the 
equation (Raychaudhuri 1955), 
 \begin{equation}
\label{e18}
{{1} \over {A}}{{d^2A} \over {ds^2}}=-R_{\alpha \beta }U^\alpha U^\beta ,
\end{equation}
where   $R_{\alpha \beta }$ is the Ricci tensor,   $U^\alpha $ is the four-velocity and, 
s is a suitable affine parameter. 
At any point the trajectory of the  geodesic bundle is tangential to the surface of a 
four-dimensional hyper-sphere with radius $r$.  
Then since the centripetal acceleration is  $c^2/r$ where   $r$ is defined by 
 \begin{equation}
\label{e19}
{{1} \over {r^2}}={{1} \over {A}}{{d^2A} \over {ds^2}}
\end{equation}
we can define   $\varepsilon $, the fractional rate of energy loss by 
\begin{equation}
\label{e20}
\varepsilon =c^2\sqrt {{{1} \over {A}}{{d^2A} \over {ds^2}}}.
\end{equation}
This relationship for   $\varepsilon $ is a function only of Riemann geometry and does 
not depend on any particular gravitational theory. 
However, Einstein's general relativity gives a particularly elegant evaluation. 
Direct application of the field equations with the stress-energy-momentum tensor    
$T_{\alpha \beta }$ gives
\begin{equation}
\label{e21}
\varepsilon =\sqrt {{{8\pi G} \over {c^2}}\left( {T_{\alpha \beta }U^\alpha 
U^\beta -\case{1}{2}Tg_{\alpha \beta }U^\alpha U^\beta } \right)},
\end{equation}
where  $T$ is the contraction of   $T_{\alpha \beta }$ and $U^\alpha$ is the four-velocity. 
Then for a gas with density  $\rho $ where the pressures are negligible the energy loss 
rate is (Crawford 1987a)
\begin{equation}
\label{e22}
\varepsilon c =-{{1} \over E}{{dE} \over {dt}}=\sqrt {{{8\pi G\left( {\rho 
c^2+p} \right)} \over {c^2}}},
\end{equation}
where x is measured along the photon's trajectory.  
This equation can be integrated to obtain
\begin{equation}
\label{e22a}
E=E_0\exp (-\varepsilon \mathop x\limits_{}).
\end{equation}
If   $\rho =n\,m_H$  and with 
(using equation \ref{e3}) 
\begin{equation}
\label{e23}
p\approx p_c=-{{\rho c^2} \over {3}}\left( {\overline {\gamma ^2}-1} \right)=-{{1} \over 
{3}}\rho c^2,
\end{equation}
then   $\varepsilon =4.54\times 10^{-27}\sqrt{n}\,\mbox{m}^{-1}$  and the predicted 
Hubble's  constant is 
\begin{equation}
\label{e24}
H=c\varepsilon =42.0\,\sqrt{n} \,\mbox{km}\cdot \mbox{s}^{-1}\cdot \mbox{Mpc}^{-1}.
\end{equation}
 With the value   $n=2.05m_H \mbox{ m}^{-3}$  we get  
$H=60.2\,\mbox{km}\cdot \mbox{s}^{-1}\cdot \mbox{Mpc}^{-1}$.  
Note for non-cosmological applications where the curvature pressure is negligible the 
results are   
$\varepsilon =5.57\times 10^{-27}\sqrt{n} \,\mbox{m}^{-1}$ or  
\begin{equation}
\label{e24a}
\varepsilon c = 51.5\sqrt{n} \,\mbox{km}\cdot \mbox{s}^{-1}\cdot \mbox{Mpc}^{-1}.
\end{equation}
Required later is the product of Hubble's constant with the radius of the universe which is  
$RH=\sqrt{2}\,c$. 
This is identical to that derived earlier (Crawford 1993) for a Newtonian cosmology.

The principle of the  focusing theorem can be illustrated by considering a very long 
cylinder of gas and Newtonian gravitation. 
At the edge of the cylinder of radius  $r$ the acceleration to-wards the center of the 
cylinder is   $\ddot r=2\pi G\rho r$ where the dots denote differentiation with 
respect to time.  
Hence  for the area  A we get   $\ddot A=4\pi G\rho A$. 
Except for the numerical constant  this is the same as that for general relativity showing 
that it is the local density that determines focusing. 
The difference of a factor of one half is because the model only includes space curvature 
and not spacetime curvature. 
In both cases distant masses have no effect. 
In particular there is no focusing and hence no energy loss in the exterior Schwartzschild 
field of a spherical mass distribution such as the sun.

Since the excitation of the photon is slowly built up along its trajectory before  the 
emission of two low energy photons any other interaction that occurs with a path length 
shorter than that between the emission of secondaries will clearly diminish their production. 
That is the excitation  can be dissipated without any extra energy loss. 
The average distance between emission of secondaries is (Crawford 1987a; 
using Heisenberg's uncertainty principle)  
 $\Delta x=\sqrt {\lambda _0/4\pi \varepsilon }$ where $h$  Plank's constant, 
 $\varepsilon $ is the fractional rate of energy loss per unit distance defined above and 
 $\lambda _0$ is the wavelength of the primary photon. 
For the cosmological plasma the secondaries would have a typical frequency of 0.02 Hz for 
a 21 cm primary and about 11 Hz for an optical photon which may be compared with the
plasma frequency of 13 Hz. Thus in most cases the secondaries will not propagate 
but will be directly absorbed by the plasma.

 The classic experiment of Pound and Snyder (1965) is an example of how the hypothesis 
of a gravitational interaction  may be tested. 
They used the Mossbauer effect to measure the energy of 14.4\,KeV (${}^{57}$Co)  gamma rays 
after they had passed up or down a 22.5\,m path in helium.  
Their result for the gravitational  redshift was in excellent agreement with the predicted  
fractional change in energy of  $2.5\times 10^{-15}$. 
The gravitational interaction theory predicts a fractional change in 
energy due to the gravitational interaction based on the density of  helium in the tube of   
 $1.25\times 10^{-12}$  which is considerably larger.  
Since their measurement was for the difference between upward and downward paths any 
effects independent of direction will cancel. 
However for these conditions although the typical path length between the emission of 
secondaries of 11m is less than the length of the apparatus it is still much longer 
than the mean free path for coherent forward scattering that is the quantum description 
of refractive index.  
In this scattering the  photon is absorbed by many electrons and after a short time delay 
(half a period) a new photon with the same energy and momentum is emitted. 
For these high energy gamma rays the binding energy of the electrons can be ignored and 
the mean free path for coherent forward scattering is given by the Ewald and Oseen 
extinction length (Jackson 1975) of   $X=1/\left( {\lambda r_0n_e} \right)$ where  
 $\lambda $ is the wavelength and   $r_0$ is the classical electron radius. 
In this case   $X=0.15\,$m that is much less than the 11\,m required for secondary emission 
and therefore the gravitational interaction energy loss will be minimal.  
The major difficulty with a laboratory test is 
in devising an experiment where   $\Delta x$ is less than the size of the apparatus and also 
less than the mean free path of any other interaction. 
Nevertheless if there are any residual effects they may be detectable  in such an experiment 
with a horizontal run using gases of different types and densities.

This inhibition of the gravitational interaction can occur in astrophysical situations. 
Consider the propagation of radiation through the Galaxy where there is a fully ionized 
plasma with density  $\rho =n\,m_H$, then the critical density is defined by when the 
Ewald and Oseen extinction length is equal to the distance between emission of secondary photons. 
If the density is greater than this critical density then the inhibition by refractive 
index impairs the gravitational interaction and there is a greatly reduced redshift. 
The critical density (for a hydrogen plasma) is   
$n_e=426.5/\lambda ^2\,\mbox{m}^{-3}$. 
For 21\,cm radiation the critical density is  $n_e=9,700\,\mbox{m}^{-3}$  and since 
most inter-stellar densities are much larger than this we do not expect 21\,cm radiation 
within the Galaxy to show redshifts due to the gravitational interaction. 
However if the gas is very clumpy we could still get uninhibited redshifts from the low
density components.
Thus all redshifts of 21\,cm radiation within the Galaxy may be  primarily due to doppler shifts.  
However optical radiation in the Galaxy should show the redshift due to the gravitational 
interaction.  This inhibition could be verified if a neutral hydrogen cloud could be
clearly identified with an object having optical line emission.

It has been argued (Zel'dovich 1963) that tired light cosmologies (such as this) should 
show a smearing out of the images of distant sources. 
The argument is that if the energy loss is caused by an interaction with inter-galactic matter, 
it is accompanied by a transfer of momentum with a corresponding  change in direction. 
That is the photon is subject to multiple scattering and hence photons from the same source 
will eventually have slightly different directions and its image will be smeared. 
For the gravitational interaction  the interaction is not with some particle with 
commensurate mass but with the mass of the gas averaged over a suitable volume. 
Since the effective mass is so large the scattering angles will be negligible. 
Furthermore in low density gas the photon loses energy to two secondary photons and to 
conserve spin and momentum these will be on average  be emitted symmetrically so that 
there is no scattering of the primary photon. Thus this model overcomes the scattering
objection to tired-light explanations for redshifts.

\section{The microwave background radiation}
Because of their wave nature electrons and other particles will be subject to the focusing 
theorem in a way similar to photons. 
In Crawford (1991) it was argued that  particles such as electrons are subject to a similar 
centripetal acceleration that produces a fractional energy loss rate of 
$\varepsilon _e$, and for a gas with density  $\rho $ and pressure $p$ it is 
 \begin{equation}
\label{e25}
\varepsilon _e=\sqrt {{{8\pi G} \over {c^2}}\left[ {(\gamma ^2 - \case{1}{2})\rho 
c^2+\left( {\gamma ^2 + \case{1}{2} } \right)p} \right]}\ ,
\end{equation}
where  $\gamma $ is the usual velocity factor. 
Hence the rate of energy loss as a function of distance  is
 \begin{equation}
\label{e26}
{{dP^0} \over {dx}}=\sqrt {{{8\pi G} \over {c^4}}\left[ {(\gamma^2-
\case{1}{2})\rho c^{^2}+\left( {\gamma ^2+\case{1}{2}} \right)p} \right]}\,\beta^2P^0,
\end{equation}
where  $\beta =v/c$ is the particle's velocity relative to the medium and $P^0$ is the energy component of its momentum four-vector. 
As it moves along its trajectory the particle will be excited  by the focusing of its wave packet.  
For charged particles conservation of spin prevents them from removing their excitation by direct emission of low energy photons. 
However if there are photons present it may interact by stimulated 
emission and thereby lose energy to secondary photons.  
The dominant photon field in inter-galactic space is that 
associated with the microwave background radiation. 
The model proposed is that the electrons lose energy by stimulated emission  to the 
background radiation so that the local black body spectrum is conserved. 
Since the conservation of energy, momentum and spin prevents a free electron from
radiating it can only lose its energy of excitation by interacting with another particle
or in this case the radiation field. 
The hypothesis is that it continuously gains energy until an interaction with a photon
stimulates the emission of a new photon. Thus the energy spectrum of the emitted photons
will match that of the existing photons. Thus give a local black body spectrum the emitted
radiation will also have the same black body spectrum. This does not explain how the
black body radiation originally arose but if there is any way in which photons can interact
to alter their energy spectrum then the equilibrium spectrum is that for a black body.

Concurrently because of the gravitational interaction the photons are losing energy that is absorbed by the plasma. 
Note that most of the secondary photons have frequencies below the plasma frequency. 
Although this means that they cannot propagate it does not prevent direct absorption of 
their energy. 
After all for frequencies below the plasma frequency the electrons can have bulk motion 
and absorb energy from an oscillating field. 
Given an equilibrium condition in which the energy lost by the electrons is equal to the 
energy lost by the photons we can equate the two energy loss rates  and get an expression 
for the temperature of the microwave background radiation (Crawford 1991) of
\begin{equation}
\label{e27}
T_M^4={{n_em_ec^3} \over {4\sigma }}\overline{ {\left( {(\gamma ^2-
\case{1}{2})+\left( {\gamma ^2+\case{1}{2}} \right){{p} \over {\rho c^{^2}}}} 
\right)\beta ^3\gamma } },
\end{equation}
where  $n_e$ is the electron number density,  $m_e$ is the electron mass,  
$\sigma $ is the Stefan-Boltzmann constant and, an average s done over all electron velocities.  
For an electron temperature of $1.11\times 10^9$\,K the bracketed term has the value 
of 0.555 with zero pressure or 0.412 with the gravitational curvature pressure.
With an electron density of 1.78\,m$^{-3}$ corresponding to a mass density of
$2.05m_H \mbox{ m}^{-3}$ and with the curvature pressure included  the predicted temperature 
is  3.0\,K. 
Given the deficiencies of the model (mainly its assumption of homogeneity) this is in good 
agreement with the observed value of 2.726\,K (Mather et al 1994). 
It is interesting that the predicted temperature only depends on the average density and a 
function of electron velocities that 
is of order one.

\section{Dark matter}
In the standard big-bang cosmology there are three major arguments (Trimble 1987) for the 
existence of dark matter, that is matter that has gravitational importance  but is not seen at any wavelength. 
The first argument is based on theoretical considerations of closure and reasonable cosmological models within the big-bang paradigm.  
The second is from the application of the virial theorem to clusters of galaxies and the third is that galactic rotation curves show high velocities at large radii. 
The first of these is purely an artifact of the big-bang cosmological model; it is not based on observation  and therefore it is not relevant to this cosmology. 
The second and third are based on observations and will be discussed at some length.

In the standard big-bang model all the galaxies in a cluster are gravitationally bound and do not partake in the universal expansion. 
If they are gravitationally bound then assuming that their differential (peculiar) redshifts are due to differential velocities we can use the virial theorem to estimate the total (gravitational) mass in the cluster. 
Typically this gravitational mass is one to several orders of magnitude larger than the mass derived from the luminosities of the galaxies: hence the need for dark matter. 

Observations of X-rays from galactic clusters show that there is a large mass of gas in the 
space between the galaxies.  
Although the mass of this inter-cluster gas is small compared to the mass of the presumed 
dark matter it is large enough to give significant redshifts due to the gravitational interaction. 
Thus the current model ascribes most of the differential redshifts to gravitational interactions 
in the inter-cluster gas. 
This model has been quantitatively investigated by Crawford (1991) for the Coma cluster. 
The method used was to take the observed differential redshift for each galaxy and by 
integrating equation (\ref{e24a}) through the known inter-galactic gas the differential 
line-of-sight distance to the galaxy was computed. 
The gas density distribution that was used is that given by Gorenstein, Huchra \& 
de Lapparent (1979). 
The result  is that galaxies with lower redshifts than that for the center of the cluster 
would be nearer and those with higher redshifts would be further away. 
The model assumed that the inter-cluster gas was spherically distributed and the test was 
in how well the distribution of Z coordinates  compared with those for the X and Y coordinates 
that were in the plane of the sky.   
Furthermore it was assumed that genuine velocities were negligible compared to the effective 
velocities of the differential redshifts. 
The median distances for each coordinate were X=0.19\,Mpc, Y=0.17\,Mpc and Z= 0.28\,Mpc. 
Given that the Coma cluster has non-spherical structure and that the model is very simple 
the agreement of the median Z distance with those for X and Y is good. 
Again it should be emphasized that there were no free parameters; the Z distances depend only on the gas distribution, the measured differential redshift, and equation (\ref{e24a}). 
If this result can be taken as representative of clusters  then there is no need for dark 
matter to explain cluster `dynamics'. 
The large differential redshifts are mainly due to gravitational interactions in the 
inter-galactic gas.

One of the difficulties with the big-bang cosmology is that it is so vague in its 
predictions that it is very difficult to refute it with observational evidence. 
However the redshifts from a cluster of galaxies can provide a critical test. 
Since celestial dynamics is time reversible a galaxy at any point in the cluster is 
equally likely to have a line-of sight velocity towards us as away from us.  
Then if accurate measurements of magnitude, size or some other variable can be used to 
get differential distances there should (in the big-bang cosmology) be no correlation 
between differential redshift and distance within the cluster.
Whereas in the static cosmology proposed here there should be a strong correlation with 
the more distant galaxies having a higher differential redshift. 
Clearly this is a difficult experiment since for the Coma cluster  it requires 
measurements of differential distances  to an accuracy of about 1\,Mpc at a distance of 100\,Mpc.

The third argument for dark matter comes from  galactic rotation curves. 
 What is observed is that  velocity plotted as a function of distance along the major 
axis shows the expected rapid rise from the center but instead of reaching a maximum and 
then declining in an approximately Keplerian manner it tends to stay near its maximum value. 
The standard explanation is that there is a halo of dark matter that extends well beyond 
the galaxy and that has a larger mass than the visible galaxy. 
For this static cosmology a partial explanation is that most of the redshift is due to 
gravitational interaction in a halo but one that is commensurate in size with the galaxy. 
Consider a spiral galaxy that is inclined to the line of sight and that has a halo 
with a Gaussian density distribution (chosen purely for analytic convenience).   
Then if the redshift origin is taken to be at the centre of the galaxy light 
from points further away will travel through more
halo gas and therefore will be redshifted and nearer points will be blue shifted.
Let the halo density distribution as a function of radius be 
$\rho = \rho_0 Exp{-(r/r_0)^2}$
and let $x$ be the distance measured from the galaxy centre along line through the galaxy
that lies in the same plane as the line of sight and the normal to the galaxy.
Then the observed redshift (in velocity units) is
\begin{equation}
\label{e27a}
v - v_0 = \pi(4G\rho_0)^{1/2} r_0 \exp(-(\frac{x \sin(i)}{\sqrt{2}r_0})^2)
erf(\frac{x \cos(i)}{\sqrt{2} r_0})
\end{equation}
where $i$ is the inclination angle.
Since the error function is asymmetric and the exponential function dominates 
at large distances the relative velocity shows a rapid increase to a broad maximum 
and then a slow decrease back to zero. 
For most galaxies it is likely that the maxima well extended
beyond the physical limits of the galaxy and so that only part of the decrease 
may be observed.
Clearly the precise shape of the curve and its numerical value will depend on the
precise nature of the density distribution. 
Nevertheless the value of the maximum velocity
for this curve will give a good indication of the effect. 
For an inclination angle of $\deg{45}$ the maximum is when $x \simeq 0.8r_0$ 
and it has the value $1.4\times 10^{-2} r_0 \sqrt{n_0}$ where $r_0$ is in kpc 
and $n_0$ is the density in H atoms per $m^3$. 
For the values $r_0 = 10$kpc and $n_0 = 10^6$ the maximum redshift
in velocity units is 140 km$s^{-1}$ which is within the range of observed values.
The difficulty with this model is that it predicts that the maximum spectral shifts 
should occur along the line of sight whereas most observations show that the
maximum velocity gradient is along the major axis.
Although it is possible to devise density distributions that can explain 
particular rotation curves there is no universal model that can explain all 
rotation curves.  
Nevertheless the fact that it predicts the magnitude and shape of a typical  
galactic redshift curve must carry some weight.

If the size and shape of the halo is a simple function of the galaxies luminosity
this model can partly explain the Tully-Fisher correlation for spiral galaxies
(Rowan-Robinson 1984) and the Faber-Jackson correlation (Faber and Jackson 1976) 
for elliptical galaxies.
They observed that the absolute magnitude of galaxies is correlated with
the width of the 21-cm emission line of neutral hydrogen.
If the line width is primarily due to gravitational interactions in the galactic
halo then its width $W_0$ is 
$W_0 = A\pi(4G\rho_0)^{1/2} r_0$ where $A$ is a constant of order unity that
depends on the actual density distribution. 
To proceed further requires knowledge of how the halo properties depend on
luminosity.

These two cases illustrate an important aspect of redshifts in this cosmology. 
Although the 
redshift is on average an excellent measure of distance any particular redshift is only a 
measure of the gas in its line of sight. 
Any lumpiness in the inter-cluster gas will produce apparent structure in redshifts that 
could be falsely interpreted as structure in galaxy distributions. 
That is, the apparent "walls", "holes", and other structures may be due to intervening 
higher density  or lower density clouds. 
For example the model predicts an apparent hole behind clusters of galaxies because of 
gravitational interactions in  intra-cluster gas. 
The velocity width of the hole would be of the same magnitude as the velocity dispersion 
in the cluster. 
For the Coma cluster the velocity width of this hole would vary from about 
4100\,kms$^{-1}$ near the center of the cluster to about 1200\,kms$^{-1}$ near the edge. 

\section{No evolution}
The most important observational difference between this cosmology and the big-bang 
cosmology is that it obeys the perfect cosmological principle: it is homogeneous both 
in space and time. 
Consequently any unequivocal evidence of evolution would be fatal to its viability.  
In contrast the big-bang theory demands evolution. 
However it has the  difficulty that the theory only provides broad guides as to what 
that evolution should be and there is little communality between the  evolution required 
for different observations.  
Nevertheless there is an entrenched view that evolution is observed in the characteristics 
of many objects. 
Two notable examples are the luminosity distribution of quasars and the angular-size 
relationship for radio galaxies. 
It will be shown that the observations for both of these phenomena are fully compatible 
with a static cosmological model. 

\section{Quasar luminosity distribution}
Because of their high redshifts quasars are excellent objects for probing the distant universe. 
Since this cosmological model is static neither the density distribution nor the luminosity 
distribution of any object should be a function of distance. 
Consider the density distribution  
$n\left(z\right)$ where z is the usual redshift parameter 
 $z=(\lambda _{\mbox{observed}}/\lambda _{\mbox{emitted}}-1$) then 
\begin{equation}
\label{e28}
z=\exp \left( {Hr/c} \right)-1,
\end{equation}
where $r$ is  the distance. 
Since the range of $r$ is  $0\le r\le \pi R$ the maximum value of z is 84.0 and its value 
at the `equator' is  8.2. 
Given that the geometry is that for a 
three-dimensional hyper-spherical surface with radius R in a four-dimensional space the 
volume out to a distance  $r$ is
\begin{equation}
\label{e29}
V\left( r \right)=2\pi R^2\left( {r-{{R} \over {2}}\sin \left( {{{2r} \over {R}}} 
\right)} \right)
\end{equation}
and the density distribution as a function of redshift for an object with a uniform 
density of   $n_0$ 
\begin{eqnarray}
\label{e30}
n\left( z \right)dz&=&n_0{{dV} \over {dr}}{{dr} \over {dz}}dz\nonumber \\
&=&{{4\pi R^2cn_0\sin ^2\left( {c\ln \left( {1+z} \right)/RH} \right)} \over {H\left( {1+z} 
\right)}}dz.
\end{eqnarray}
From equations (\ref{e14}) and (\ref{e24}) we find that  $HR=\sqrt 2\,c$ and equation 
(\ref{e30}) becomes 
 \begin{equation}
\label{e31}
n\left( z \right)dz={{4\pi R^3n_0\sin ^2\left( {\ln \left( {1+z} \right)/\sqrt 2} 
\right)} \over {\sqrt 2\left( {1+z} \right)}}dz,
\end{equation}
which has a maximum when  $z=2.861$. 
Now the difficulty of using equation (\ref{e31}) with observations is that most quasar 
observations  have severe selection effects. 
Boyle et al (1990) measured the spectra of 1400 objects of which 351 were identified 
as quasars with redshifts  z $<$ 2.2. 
The advantage of their observations is that their selection effects were well defined. 
A full analysis is given in Crawford (1995b). 

Let a source have a luminosity  $L\left( \nu \right)$(Whz$^{-1}$) at the emission 
frequency $\nu$. 
Then if the energy is conserved the observed flux density 
 $S\left( \nu  \right)$ (Wm$^{-2}$Hz$^{-1}$) at a distance  $r$  is the luminosity 
divided by the area which is
\begin{equation}
\label{e32}
S\left( \nu  \right)={{L\left( \nu  \right)} \over {4\pi R^2\sin ^2\left( {r/R} 
\right)}}.
\end{equation}
However because of the gravitational interaction there is an energy loss such that the 
received frequency   $\nu _0$ is related to the emitted frequency   $\nu _e$ by 
\begin{equation}
\label{e33}
\nu _0=\nu _e\exp \left( {-Hr/c} \right)=\nu _e/\left( {1+z} \right).
\end{equation}
This  loss in energy means that the observed flux density is decreased by a factor of  $1+z$. 
But there is an additional bandwidth factor  that exactly balances the energy loss
factor. 
In addition allowance must be made for K-correction (Rowan-Robinson 1985) 
that relates 
the observed spectrum to the emitted spectrum. 
Since it is usual to include the bandwidth factor in the K-correction the apparent magnitude is
 \begin{eqnarray*}
\label{e34}
m & =&  -\case{5}{2}\log \left( {S\left( {\nu _0} \right)} \right)\nonumber \\
  & =&  -\case{5}{2}\log \left( {L\left( {\nu _0} \right)} \right)+
\case{5}{2}\log \left( {4\pi R^2} \right)
\nonumber \\
&+&
5\log \left( {\sin \left( {{{c\ln (1+z)} 
\over {HR}}} \right)} \right) \\
&+&\case{5}{2}\log \left( {1+z} \right)
+K\left( z \right),
\end{eqnarray*}
where  $K\left( z \right)$ is the K-correction and from above $RH=\sqrt{2}c$.
The result of the analysis was that the observations were fitted by a (differential)  
luminosity function  that had a Gaussian shape with a standard deviation of 1.52 magnitudes 
and a maximum at  $M=-22.2\,\mbox{mag}$ (blue). 
The only caveat was that there appeared to be a deficiency of weak nearby quasars in the sample. 
Since all cosmological models are locally Euclidean this must be a selection effect.
The fact that the absolute magnitude distribution had a well-defined peak and this was achieved 
without requiring any  evolution is strong support for the static model.

\section{Angular size of radio sources}
For the geometry of the hyper-sphere the observed angular size   $\theta $ for an object with 
a redshift of  z and projected linear size of D is   
$\theta =D/\left( {R\sin \left( {r/R} \right)} \right)$, and in terms of redshift it is 
 \begin{eqnarray*}
\label{e35}
\theta &=&{D \over {R\sin \left( {c\ln \left( {1+z} \right)/RH} \right)}}\\
&=&{D \over {R\sin \left( {\ln \left( {1+z} \right)/\sqrt 2} \right)}}.
\end{eqnarray*}
The angular size decreases with z until   $z=8.2$ where there is a broad minimum and then 
it increases again. 
This model was used (Crawford 1995a) to analyze 540 double radio sources (all 
Faranoff-Riley type II) listed by Nilsson et al (1993). 
The result was an excellent fit to the radio-source size measurements, much better than 
the big-bang model with a free choice of its acceleration parameter.  

\section{Surface brightness of galaxies}
In an expanding universe, bolometric surface brightness  will decrease with redshift as 
$(1+z)^{-4}$ while, in a nonexpanding cosmology with tired-light it will decrease as
$(1+z)^{-1}$.
Because it is independent of the geometry of space it is a powerful test for discriminating
between the two cosmologies. The problem is that the measurement of surface brightness
is very difficult. Recently Pahre et al (1996) have reported measurements of surface
brightness for elliptic galaxies in the clusters Abell 2380 ($z=0.23$) and Abell 851 ($z=0.41$)
and have compared them with the nearby Coma cluster. 
Their final results are surface brightness measurements for an average elliptic galaxy in
the K, B and $R_c$ spectral bands for the Coma and Abell 851 clusters and only the K band
for cluster Abell 2380. 
Although they claim that the results are inconsistent with a tired light cosmology
their claim is only true for the K band data. 
The B and $R_c$ band data are consistent with a tired light cosmology. 
In addition the K-corrections for the K band data seem to have been computed using
evolving galaxy models whereas for a valid comparison it should be done for a 
non-evolving galaxy. 
Although the difference is small the K corrections are commensurate with the
discrepancy with the tired light model.
It would be more convincing if the K band observations for Abell 851 could be done
with a filter redshifted by a factor of 1.41.

\section{Other evidence for evolution}
There are however more direct observations of evolution that will be discussed. 
They are the 
distribution of absorption lines in quasar spectra, the measurement of the 
microwave background temperature at high redshift, the time dilation of the type I 
supernova light curves at large distances and the Butcher-Oemler effect.
For this static cosmology consider a uniform distribution of objects with number density N and cross-sectional area A  then their distribution in redshift along
 a line of sight is (here   $\gamma $ is the exponent and not the Lorentz velocity parameter)
\begin{equation}
\label{e36}
{{dN} \over {dz}}={{NAc} \over {H}}\left( {1+z} \right)^\gamma \ .
\end{equation}
with   $\gamma =-1$. 
If the  absorption lines seen in the spectra of quasars are due to absorption by 
the Lyman-$\alpha$ line of hydrogen in intervening clouds of gas and with a 
uniform distribution of clouds their predicted redshift  distribution should 
have   $\gamma =-1$. 
However observations (Hunstead et al 1988; Morris et al 1991; Williger et al 1994; 
Storrie-Lombardie et al 1997; Barlow \& Tytler 1998) show exponents that range from 0.8 to 4.6. 
Although there is poor agreement amongst the observations clearly they are all 
in disagreement with this model. 
The recent observations of Barlow and Tytler (1998) are of interest in that for the 
Lyman-$\alpha$ lines they get $\gamma \simeq 1$ but for C IV $\lambda1548.20$ absorption lines
they  find that the result from Steidel (1990) of $\gamma = -2.35\pm 0.77$ is inconsistent
with their low z data point and that equation \ref{e36} has a very poor fit.

Observations of absorption lines have complications due to lack of resolution causing 
lines to be merged and that only a limited range in z (from Lyman-$\alpha$ to 
Lyman-$\beta$)  is available from each quasar. 
However the major change required in the interpretation of the results  for the static 
cosmology  is in the explanation for the broad absorption lines. 
Traditionally these have been ascribed to Doppler 
broadening from bulk motions in the clouds but it is also possible that they are due to 
energy loss by the gravitational interaction. 
For example using equation (\ref{e24a}) the `velocity' width of a cloud of diameter 
$10^4$\,pc and density  $10\,m_Hm^{-3}$ is  $16\,\mbox{km}\cdot \mbox{s}^{-1}$ which is 
typical of the observed line widths. 
For a typical column density of  
$ N_{\mbox{H1}}=10^{15}\,\mbox{cm}^{-2}$ this 
cloud would have a  ratio of H$_1$ to ionized hydrogen of  
$3\times 10^{-5}$. 
A further consequence is that because of the clouds the observed redshift 
is not a valid measure of the true distance. 
For example suppose the quasar is located in a galactic cluster where we would 
expect a high local concentration of clouds then its redshift would be increased 
over that expected for the cluster by the extra energy loss in the clouds. 
The conclusion is that until the nature if the absorption lines are better 
understood and analyzed in the context of this theory the evidence for 
evolution is not convincing.

Another observation that could refute this theory is if the cosmic microwave background 
radiation has a higher temperature at large distances. 
Ge et al (1997) measured the absorption from the ground and excited states 
of C1 (with a redshift of 1.9731)  in the quasar  QSO 0013-004. 
They measure the strengths of the J=0 and J=1 fine structure levels  and derived an 
excitation temperature of   $11.6\pm 1.0$\,K which after corrections gives a temperature 
for the surrounding radiation of  $7.9\pm 1.0$\,K that is in good agreement with the 
redshifted temperature of 8.1\,K. 
On face value this is clear evidence for evolution. 
But not only are the measurements difficult they are based on a model for line widths 
that does not include broadening due to the gravitational interaction. 
Until this is done and the results are confirmed for other quasars and by other 
observers a static cosmology is not refuted. 

Programs that search for supernovae in high redshift galaxies with large telescopes are now 
finding many examples and more importantly some are being detected before they reach their 
maximum intensity. 
Leibundgut et al (1996),  Goldhaber et al (1996), and Riess et al (1997) have reported on 
type 1a supernovae in which they believe that they can identify the type of supernova from 
its spectral response and by comparing the supernova light curves with reference templates 
they measure a time dilation that corresponds to that expected for their redshift in a 
big-bang cosmology.  
In addition there is a significant correlation between the rate of decay of the light curve
and the intrinsic luminosity (Riess, Press and Kirshner 1996) in that brighter supernovae
have a slower decline.
Hence there may be a bias due to selection effects and the cosmological model used
to get the absolute luminosity that is needed to correct for the correlation.
However because of this correlation and of uncertainties in matching the exact type 
of supernova and because of the 
occurrence of individual inhomogeneities many more observations are needed before these results 
are well established.

The Butcher-Oemler (1978) effect is the observation that galaxies at redshifts $z \geq 0.3$ 
the galaxies in rich clusters tend to be bluer than is typical of nearby clusters.
Couch et al (1998) have found significant differences in their study of three rich clusters
at a redshift of $z=0.31$. 
However Andreon and Ettori (1999) looked at a larger sample of x-ray selected clusters and
found no evidence for the Butcher-Oemler effect.
Their argument is that the effects that are observed are due to selection criteria rather 
than differences in look back time.
As well as lack of unambiguous observations the effect (when present) is only seen at redshifts 
up to $z \sim 1$ which is only relevant to the local inhomogeneity.

The conclusion is that the Lyman-$\alpha$ forest observations and the cosmic background radiation 
temperature observations need to be re-evaluated within  the static cosmological model in 
order to see if they show evolution and refute the model. 
The supernovae results are essentially unchanged in the static model and if they hold up they 
make a strong case for evolution that would refute any static model.
The Butcher-Oemler effect observations are still not strong enough to make a good argument
against a static homogeneous universe.

\section{Nuclear abundance}
In this cosmology the universe is dominated by a very high temperature plasma. 
Galaxies 
condense from this plasma, evolve and die. 
Eventually all of their matter is returned to the plasma.  
Although nuclear synthesis in stars and supernova can produce the heavy elements it cannot 
produce the very  light elements. 
In big-bang cosmology these are produced  early in the expansion when there were high 
temperatures and a large number of free neutrons. 
This mechanism is not available in a static cosmology. 
Nevertheless the temperature of the plasma  ($2\times 10^9\,\mbox{K}$)  is high enough 
to sustain nuclear reactions. 
The end result is an abundance distribution dominated by hydrogen and with smaller 
quantities of helium and other light elements. 
The problem is that the density is so low that the reaction rates may be too small achieve 
equilibrium within the recycling time. 
Naturally much further work is needed to quantify this hypothesis.

\section{Entropy}
Nearly every textbook  on elementary physics quotes a proof based on  the second law of 
thermodynamics to show that the entropy of the universe is increasing but this is in direct 
conflict with the perfect cosmological principle  where total entropy is constant. 
The conflict can be resolved if it is noted that the formal proof of the second law of 
thermodynamics requires consideration of an isolated system and the changes that occur 
with reversible and irreversible heat flows between it and its surroundings. 
Now there is no doubt that irreversible heat flows occur and lead to an overall increase in 
entropy. 
However the formal proof is flawed in that with gravitational fields one cannot have an 
isolated system. 
There is no way to shield gravity. 
Furthermore in their delightful  book Fang \& Li (1989) argue that a self-gravitating system 
has negative thermal capacity and that such systems cannot be in thermal equilibrium.  
The crux of their argument is that if energy is added to a self-gravitating system, such as the 
solar system, then the velocities and hence the `temperature' of the bodies decrease. 
What happens is that from the virial theorem the potential energy (with a zero value for a 
fully dispersed system)  is equal to minus twice the kinetic energy and the total energy is 
the sum of the potential and kinetic energies which is therefore equal to minus the kinetic 
energy. 
Thus we must be very careful in applying simple thermodynamic laws to gravitational systems. 

Now consider the gravitational interaction  where photons lose energy to the background 
plasma. 
Since this process does not depend on temperature it is not a flow of heat energy rather it 
is work. 
Nevertheless we can compute the entropy loss from the radiation field, considered as a heat 
reservoir, as  $-W/T_r$ where W is the energy lost, and similarly the entropy gained by the 
plasma as  $W/T_e$. 
Then since  $T_e>>T_R$ there is a net entropy loss. 
Thus this gravitational interaction not only produces the Hubble redshift but it also acts 
to decrease the entropy of the universe thereby balancing the entropy gained in irreversible 
processes such as the complementary interaction where electrons lose energy to the radiation 
field.  

\section{Olber's paradox}
An essential requirement of any cosmology is to be able to explain Olber's paradox (or more 
correctly de Chesaux's paradox; Harrison 1981) as to why the sky is dark at night. 
For the big-bang cosmology although the paradox is partly explained by the universal redshift 
the major  reason is that the universe has a finite lifetime. 
For this static cosmology the explanation is entirely due to the redshift. 
The further we look  to distant objects the more the light is redshifted until it is shifted 
outside our spectral window. 
Thus in effect we only see light from a finite region.  
Note that the energy lost by the photons is returned to the inter-galactic plasma as part of 
a cyclic process. 

\section{Conclusion}
The introduction of curvature pressure has wide ranging astrophysical applications. It is 
possible that it may resolve the solar neutrino problem but this must await a full analysis 
using the standard solar model.  
Although the theory does not prevent the formation of a black hole from cold matter it does 
have an important effect on the high temperature accretion rings and curvature pressure 
may provide the engine that produces astrophysical  jets. 

The greatest strength of this model is that it shows how a stable and static cosmology 
may exist within the framework of general relativity without a cosmological constant.  
The model with a homogeneous plasma depends only on one parameter, the average density 
which from X-ray observations is taken to be  $2.05m_H \mbox{ m}^{-3}$. 
It then predicts that the plasma has a temperature of  $2\times 10^9\,K$ and that the 
universe has a radius given by equation (\ref{e14}).  
It has been shown that for a simple inhomogeneous density distribution the predicted 
temperature could easily be much lower and it could be in agreement with the temperature 
observed for the X-ray background radiation.  
Inclusion of the gravitational interactions permits the prediction of a Hubble constant 
of   $H=60.2\,\mbox{km}\cdot \mbox{s}^{-1}\cdot \mbox{Mpc}^{-1}$ and a microwave 
background radiation with a temperature of  $3.0\,K$.  
Dark matter does not exist but arises from assuming that non-cosmological redshifts are 
genuine velocities and then using  the virial theorem. 
In this static model most of the non-cosmological  velocities are due to gravitational 
interactions in intervening  clouds. 

Analysis of the observations for quasar luminosities and  the angular size of radio sources 
shows that they can be fully explained in a static cosmology without requiring any evolution. 
The implication is that many other observations that require evolution in the big-bang 
cosmology need to be re-examined within the static cosmology before evolution can be confirmed.  
The strong evolution shown in the distribution of absorption lines (the Lyman-$\alpha$ forest) 
is a problem for the static model. 
However because of the gravitational interaction that can cause line broadening and the 
possibility that some of the redshift may come from the clouds that produce the absorption 
lines the results cannot at this stage be taken as a refutation of the static model.
Although the observations of a redshifted background microwave temperature and the evidence 
of time dilation in the decay curves of type 1a supernovae appear to show direct evolution  
it is too early to be certain.  
These observations need  better statistics and should be analyzed within this static model 
before their apparent evolution is convincing.  
The model includes a qualitative model for the generation of the light elements in the 
high temperature inter-galactic plasma. 
It was also argued that the effects of gravitational interaction of the microwave background 
radiation that transfers energy to the high temperature plasma decreases entropy so that overall 
total entropy of the universe is constant. 
Finally the sky is dark at night because the light from distant stars is redshifted out of our 
spectral window.  

An important characteristic of this static cosmology is that it is easily refuted: any 
unequivocal evidence for evolution would disprove the model. 
Apart from evolution the most discriminating test that chooses between it and the big-bang 
cosmology would be to compare the differential velocities of galaxies in a cluster with 
their distance. 
Whereas the big-bang model  requires that there is no correlation this static cosmology requires 
that the more distant galaxies will have larger redshifts.

\section{Acknowledgments}
I wish the thank the referee for many critical comments of the original draft 
and for suggesting several recent references. 
This work is supported by the Science Foundation for Physics within the University of Sydney, 
and use has made of NASA's Astrophysics Data System Abstract Service. 

\section{References}
\setlength{\leftmargin=1em}
\setlength{\itemindent=-\leftmargin}

\reference{1} Allen, C. W. 1976, Astrophysical Quantities, 3rd ed. (Athlone; London)
\reference{1a} Andreon, S., \& Ettori, S., 1999, \apj, {\bf 516}, 647.
\reference{2} Bahcall, J. 1989, Neutrino Astrophysics, (Cambridge University Press; Cambridge).
\reference{3} Bahcall, J. 1997, Proc. of the 18th Texas Symposium on Relativistic Astrophysics, 
eds, A. Olinto, J. Frieman, \& D. Schramm (World Scientific; Singapore).
\reference{4} Barlow, T. A. \& Tytler, D. 1998, \aj, {\bf 115}, 1725.
\reference{4a} Braginsky, V. B. \& Panov, V. I., 1971, {\it Zh. Eksp. \& Teor. Fiz.}, 
{\bf 61}, 873.
\reference{4a} Boyle, B. J., Fong, R., Shanks, T., \& Peterson, B. A., 1987, \mnras, 
{\bf 227}, 717.
\reference{5} Butcher, H., \& Oemler, A., 1978, \apj, {\bf 219}, 18.
\reference{8} Crawford, D. F. 1987a, Australian J. Phys., {\bf 40}, 449.
\reference{9} Crawford, D. F. 1987b, Australian J. Phys., {\bf 40}, 459.
\reference{10} Crawford, D. F. 1991, \apj, {\bf 377},  1.
\reference{11} Crawford, D. F. 1993, \apj, {\bf 410}, 488 .
\reference{12} Crawford, D. F. 1995a, \apj, {\bf 440}, 466.
\reference{13} Crawford, D. F. 1995b, \apj, {\bf 441}, 488.
\reference{13a} Faber, S. M. \& Jackson, R. E., 1976, \apj, {\bf 204}, 668.
\reference{14} Fabian, A. C. \&  Barcons, X. 1992, \araa, {\bf 30}, 429.
\reference{15} Fang, Li-Zhi, \& Li, Shu Xian 1989, Creation of the Universe, 
(World Scientific; Singapore).
\reference{16} Harrison, E. R. 1981, Cosmology, (Cambridge University Press; Cambridge).
\reference{17} Hunstead, R. W., Murdoch, H. S., Pettini, M., \& Blades, J. C. 1988, \apj, 
{\bf 329}, 527.
\reference{18} Ge, J., Bechtold, J., \& Black, J. H. 1997, \apj, {\bf 474}, 67.
\reference{19} Goldhaber, G., Deustua, S., Gabi, S>, Groom, D., Hook, I., Kim, A.,
Kim, M., Lee, J., Pain, R., Pennypacker, C>, Perlmuter, S., Small, I., Goobar, A.,
Ellis, R., McMahon, R., Boyle, B., Bunclark, P., Carter, D., Glazebrook, K., Irwin, M., 
Newberg, H., Filippenko, A. V., Matheson, T., Dopita, M., Mould, J., \& Couch, W,
1996, Thermonuclear Supernovae (NATO ASI), eds., 
R. Canal, P. Ruiz-Lapuente \& J. Isern (NATO ASI Ser. C, 486),(Kluwer: Dordrecht).
\reference{21} Gorenstein, P., Huchra, J. P., \& de Lapparent, V. 1979, in IAU Symposium 124, 
\reference{22} de Groot, S. R. ,Leeuwen, W. A.,\& van Weert, C. G. 1980, 
Relativistic Kinetic Theory (North-Holland; Amsterdam).
\reference{23} Jackson, J. D. 1975, Classical Electrodynamics, (John Wiley; New York).
\reference{23a} LaViolette, P. A., 1986, \apj, {\bf 301}, 544.
\reference{25} Leibundgut, B.,et al., R. 1996, \apjl, {\bf 466}, L21.
\reference{26} Mather, J. C., Cheng, E. S., Cottingham, D. A., Eplee Jr, R. E., 
Fixsen, D. J., Hewagama, T., Isaacman, R. B., Jensen, K. A., Meyer, S. S., 
Noerdlinger, P. D., Read, S. M., Rosen, L. P., Shafer, R. A., Wright, E. L.,
Bennett, C. I., Boggess, N. W., Hauser, M. G., Kelsall, T., Moseley Jr, S. H., 
Silverberg, R. F., Smoot, G. F., Weiss, R., \& Wilkson, D. T., 1994, \apj, {\bf 420}, 439.
\reference{27} Misner, C. W., Thorne, K. S., \& Wheeler, J. A. 1973, Gravitation, 
(Freeman; San Francisco).
\reference{28} Nilsson, K., Valtonen, M. J., Kotilainen, J., \& Jaak\-kola, T. 1993, \apj, 
{\bf 413}, 453.
\reference{28a} North, J. D., "The Norton History of Astronomy and Cosmology",
1995, (W. W. Norton \& Co, N.Y.).
\reference{28a} Peebles, P. J., 1993, 'Principles of Physical Cosmology', (Dover: New York).
\reference{29} Pound, R. V., \& Snider, J. L. 1965, \prb {\bf 140}, 788.
\reference{30} Raychaudhuri, A. K. 1955, Phys. Rev., {\bf 98}, 1123.
\reference{31} Reber, G., 1982, Proc ASA {\bf 4}, 482.
\reference{32a} Riess, A. G., s, W. RH., Kirshner, R. P., 1996, \apj, {\bf 473}, 88.
\reference{32} Riess, A. G., et al., R. C. 1997, \aj, {\bf 114}, 722.
\reference{33} Roll, P. G., Krotov, R., \& Dicke, R. H. 1964, Annals of Physics, {\bf 26}, 442.
\reference{33a}  Rowan-Robinson, M, 1984, "The Cosmological Distance Ladder", 
(W. H. Freeman).
\reference{33b} Steidel, C. C., 1990, \apjs, {\bf 72}, 1.
\reference{34} Storrie-Lombardi, L. J., McMahon, R. G., Irwin, M. J., \& Hazard, C. 1997, 
ESO Workshop on QSO Absorption Lines, \apj, {\bf 468}, 121.
\reference{35} Trimble, V. 1987, \araa, {\bf 25}, 425.
\reference{36} Weinberg, S. 1972, Gravitation and Cosmology (Wiley; New York).
\reference{37} Williger, G. M., Baldwin, J. A., Carswell, R. F., Cooke, A. J., Hazard, C., 
Irwin, M. J., McMahon, R. G., \& Storrie-Lombardi, L. J. 1994, \apj, {\bf 428}, 574.

\end{document}